\documentclass[12pt,a4paper]{article}
\usepackage{epsfig}
\begin{document}
\noindent \thispagestyle{empty}
\begin{center}
{\bf Sensitivity of Cross Sections for Elastic Nucleus-Nucleus
Scattering to Halo Nucleus Density Distributions} \\

\vspace{5mm}
G.~D.~Alkhazov, V.~V.~Sarantsev
\vspace{5mm}

{\it Petersburg Nuclear Physics Institute, Gatchina, RUSSIA}
\end{center}
\vspace{10mm}
\par
In order to clear up the sensitivity of the nucleus-nucleus
scattering to the nuclear matter distributions of exotic halo
nuclei, we have calculated differential cross sections for elastic
scattering of the $^6$He and $^{11}$Li nuclei on several nuclear
targets at the energy of 0.8 GeV/nucleon with different assumed
nuclear density distributions in $^6$He and $^{11}$Li.
\vspace{5mm}
\par
\noindent {\bf Comments:}14 pages, 7 figures. Submitted to Proceedings of the 
61 International Conference "Nucleus-2011" on the Problems of the 
Nuclear Spectroscopy and the Atomic Nuclear Structure, Sarov 
Nijzegorodskaya district, October 10-14, 2011.\\
\par
\noindent {\bf Report:} 
Petersburg Nuclear Physics Institute preprint 
PNPI-2874,10(2011) \\
\par
\noindent {\bf Category:} Nuclear Theory(nucl-th); High-Energy 
Physics--Theory (hep-th). \\
\par
\section{Introduction}
As has been shown in [1,2], the elastic proton-nucleus scattering
at intermediate energy in inverse kinematics is an efficient means
of studying nuclear matter density distributions in exotic halo
nuclei. An analysis of the measured cross sections for
small-angular elastic proton-nucleus scattering allows one to
determine both the sizes of the nuclear core and the neutron halo.
At the same time, as has been shown in [1], the sensitivity of the
cross sections for elastic proton-nucleus scattering is not
sufficient for obtaining information on the  matter distribution
of the nuclear far periphery, which contains small amount of the
nuclear matter. The authors of paper [3] discuss the sensitivity
of the reaction cross sections (that is of the integrated cross
sections for all inelastic processes) to the nuclear density
distribution at large distances from the nuclear centre. They come
to the conclusion that the reaction cross sections for
nucleus-nucleus scattering are significantly more sensitive to the
nuclear periphery than the reaction cross sections for
proton-nucleus scattering. The results of the paper allow us to
suppose that the cross sections for elastic nucleus-nucleus
scattering are also more sensitive to the nuclear periphery than
the cross sections for elastic proton-nucleus scattering. If it is
really so, then it is reasonable to perform the relevant
experiments, for example, at the future nuclear facility FAIR at
Darmstadt.
\section{Density distributions}
In the present work, in order to find out the sensitivity of the
cross sections for nucleus-nucleus elastic scattering to the nuclear
periphery, we have performed calculations of the cross sections for
elastic scattering of exotic nuclei $^6$He and $^{11}$Li on protons
and nuclear targets $^4$He, $^9$Be, $^{12}$C, $^{58}$Ni, $^{90}$Zr
and $^{208}$Pb at the 800 MeV/nucleon energy of the incident nuclei,
different matter distributions being assumed in the $^6$He and
$^{11}$Li nuclei. Theoretical and experimental investigations have
shown that the cross sections for proton-nucleus scattering at the
intermediate energy 
( $\sim$0.5--1.0 GeV/nucleon) can be calculated
fairly accurately with the help of the Glauber theory [4]. It should
be noted however that calculations of the cross sections for
nucleus-nucleus scattering with the exact formula of the Glauber
theory is a rather complicated task. In this work, the cross
sections for nucleus-nucleus scattering were calculated with the
Glauber theory within the "rigid-target" approximation [5], that is
at first the amplitude of scattering of one nucleon on the nuclear
target was calculated, and then this amplitude was used in the
calculations of the scattering of an exotic nucleus consisting of
several nucleons. As was shown in [6], the reaction cross sections
for scattering of exotic nuclei on nuclear targets calculated within
the "rigid-target" approximation are very close to those calculated
with the exact Glauber formula.

In the present calculations it was assumed that the $^6$He and
$^{11}$Li nuclei consist of the nuclear core of 4 and 9 nucleons,
correspondingly in $^6$He and $^{11}$Li, and 2 halo neutrons. The
matter density distributions in the core were described by Gaussian
distributions, whereas the density distributions in the halo were
described by a Gaussian for $^6$He and by a 1p-shell harmonic
oscillator-type function for $^{11}$Li. Spin-independent
isospin-avera- ged amplitude of the free nucleon-nucleon (NN)
scattering was used, the traditional high-energy parametrization for
this amplitude being taken (see [1]) with the following parameters:
the total cross section $\sigma_{\rm NN}$ = 42.5 mb, the ratio of
the real to imaginary part $\epsilon_{\rm NN} =-$0.18, and the
amplitude slope $\beta_{\rm NN}$ = 0.2 fm$^2$.

Figures 1 and 2  show the $^6$He matter density distributions
applied in the calculations. The solid curve in Fig. 1 corresponds
to the nuclear density distribution with a halo: R$_{\rm c}$ =
1.95 fm, 
R$_{\rm h}$ = 2.88 fm and R$_{\rm m}$ = 2.30 fm, where
R$_{\rm c}$, R$_{\rm h}$ and R$_{\rm m}$ are the rms radii of the
core, halo and total matter density.
\begin{figure}[h]
\centering \epsfig{file=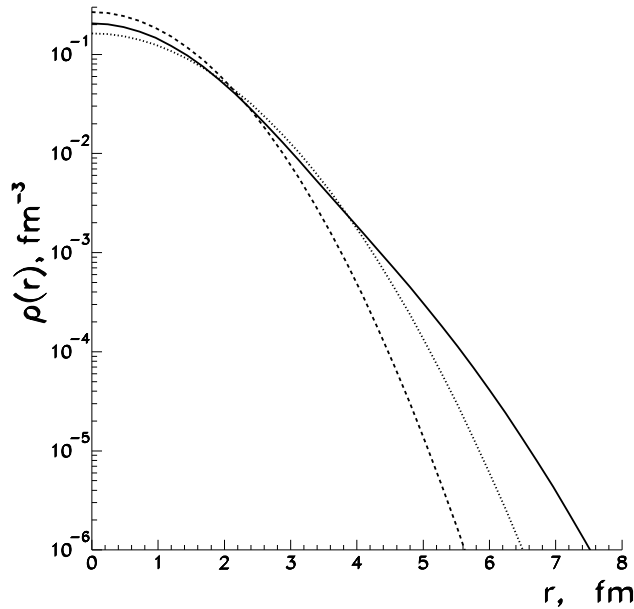,height=8cm}
\caption{\small The nuclear matter density distributions in $^6$He
(with a halo structure -- solid curve, and without a halo --
dashed and dotted curves) applied in the cross sections
calculations (for the parameters of the density distributions see
the text)}
\end{figure}
The dashed and dotted curves in Fig. 1 correspond to the density
distributions without a halo (R$_{\rm c}$ = R$_{\rm h}$), the dashed
curve stands for R$_{\rm m}$ = 1.95 fm, and the dotted curve stands
for  R$_{\rm m}$ = 2.30 fm.

The solid curve in Fig. 2
corresponds to the solid curve in Fig. 1. The halo density
distribution in this version of the calculations, as it has been
already said, is described with a Gaussian distribution. The
density in this distribution decreases with increasing the
distance $r$ from the nuclear centre faster than it is predicted
by theory.
\begin{figure}[h]
\centering \epsfig{file=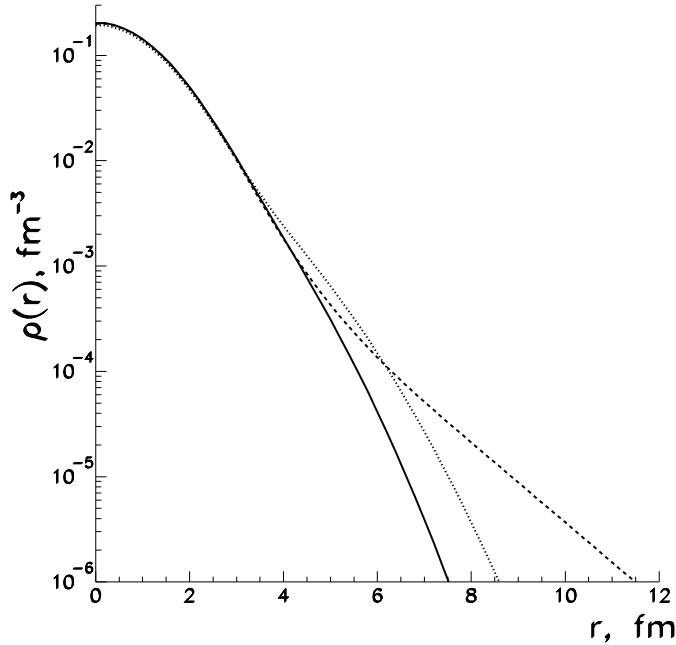,height=8.5cm}
\caption{\small The nuclear matter density distributions in $^6$He
applied in the cross sections calculations, with a density tail --
dashed curve, and without a density tail -- solid and dotted
curves}
\end{figure}
The dashed curve in Fig. 2 shows the $^6$He density distribution
where we have added a "tail", which decreases exponentially with
the radius $r$ increasing. The resulting density distribution with
this tail corresponds to the theoretical nuclear matter
distribution FC of Ref. [7]. The halo rms radius in this case is
equal to R$_{\rm h}'$ = 3.34 fm. Figure 2 shows also by the dotted
curve the $^6$He density distribution without a tail, but with the
increased halo radius: R$_{\rm h}$ = R$_{\rm h}'$ = 3.34 fm.

Figure 3 shows the $^{11}$Li density distributions used in the
calculations.
\begin{figure}[h]
\centering \epsfig{file=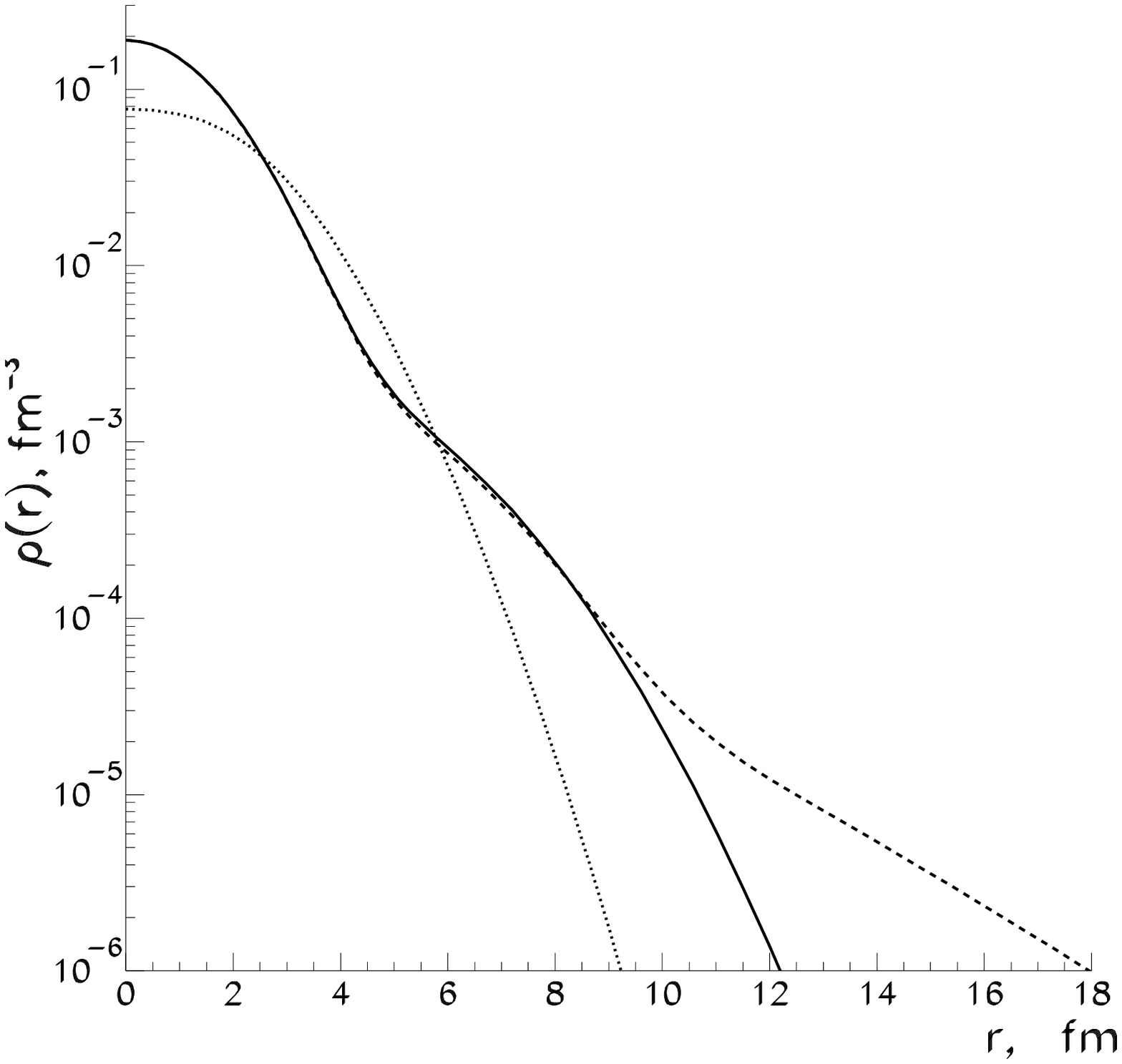,height=8.5cm} \caption{\small
The nuclear matter density distributions in $^{11}$Li applied in
the cross sections calculations (for the parameters of the density
distributions see the text)}
\end{figure}
The solid curve in Fig. 3 corresponds to the density distribution
with a halo (however without a tail): R$_{\rm c}$ = 2.50 fm,
R$_{\rm h}$ = 5.86 fm, R$_{\rm m}$ = 3.37 fm. The dotted curve
shows the density distribution without a halo (R$_{\rm c}$ =
R$_{\rm h}$, R$_{\rm m}$ = 3.37 fm), and the dashed curve shows
the density distribution with a halo and the tail corresponding to
the theoretical density distribution P2 of Ref. [8].

The matter
density distribution of the $^4$He target nucleus was described by
a Gaussian distribution with the rms matter radius R$_{\rm m}$ =
1.49 fm [1]. For heavier nuclear targets, the Fermi distribution
\vspace{-1mm}
\begin{equation}
\rho(r)\sim [1 + {\rm w}(r/{\rm R_0})]/\{1 + {\rm exp}[(r - {\rm
R_0)/a}]\},
\end{equation}
was used with the following parameters: \vspace{4mm} \\R$_0$ =
1.95 fm, a = 0.60 fm, w = 0 (R$_{\rm m}$ = 2.68 fm) for $^9$Be
[9];\\R$_0$ = 2.12 fm, a = 0.52 fm, w = 0 (R$_{\rm m}$ = 2.52 fm)
for $^{12}$C [9];\\R$_0$ = 3.17 fm, a = 0.59 fm, w = $-$0.12
(R$_{\rm m}$ = 3.15 fm) for $^{28}$Si [10];\\R$_0$ = 4.23 fm, a =
0.55 fm, w = $-$0.13 (R$_{\rm m}$ = 3.76 fm) for $^{58}$Ni
[11];\\R$_0$ = 4.99 fm, a = 0.57 fm, w = $-$0.09 (R$_{\rm m}$ =
4.35 fm) for $^{90}$Zr [12];\\R$_0$ = 6.70 fm, a = 0.54
fm, w = $-$0.06 (R$_{\rm m}$ = 5.54 fm) for $^{208}$Pb [12].\\
(Note that the used Fermi distributions include the final size of
the nucleon). When performing the calculations of the cross
sections only the centre-of-mass correlations were taken into
account in the many-body density distributions (see [1]). The
results of the calculations are presented in Figs. 4--7.
\section{Results and discussion}
 Figure 4 shows the calculated cross sections for elastic $^6$He-$p$,
$^6$He-$^4$He, $^6$He-$^{12}$C and $^6$He-$^{58}$Ni scattering.
The solid curves correspond to the calculations with the halo
density distributions in $^6$He: R$_{\rm c}$ = 1.95 fm, R$_{\rm h}$
\begin{figure}[h]
\centering \epsfig{file=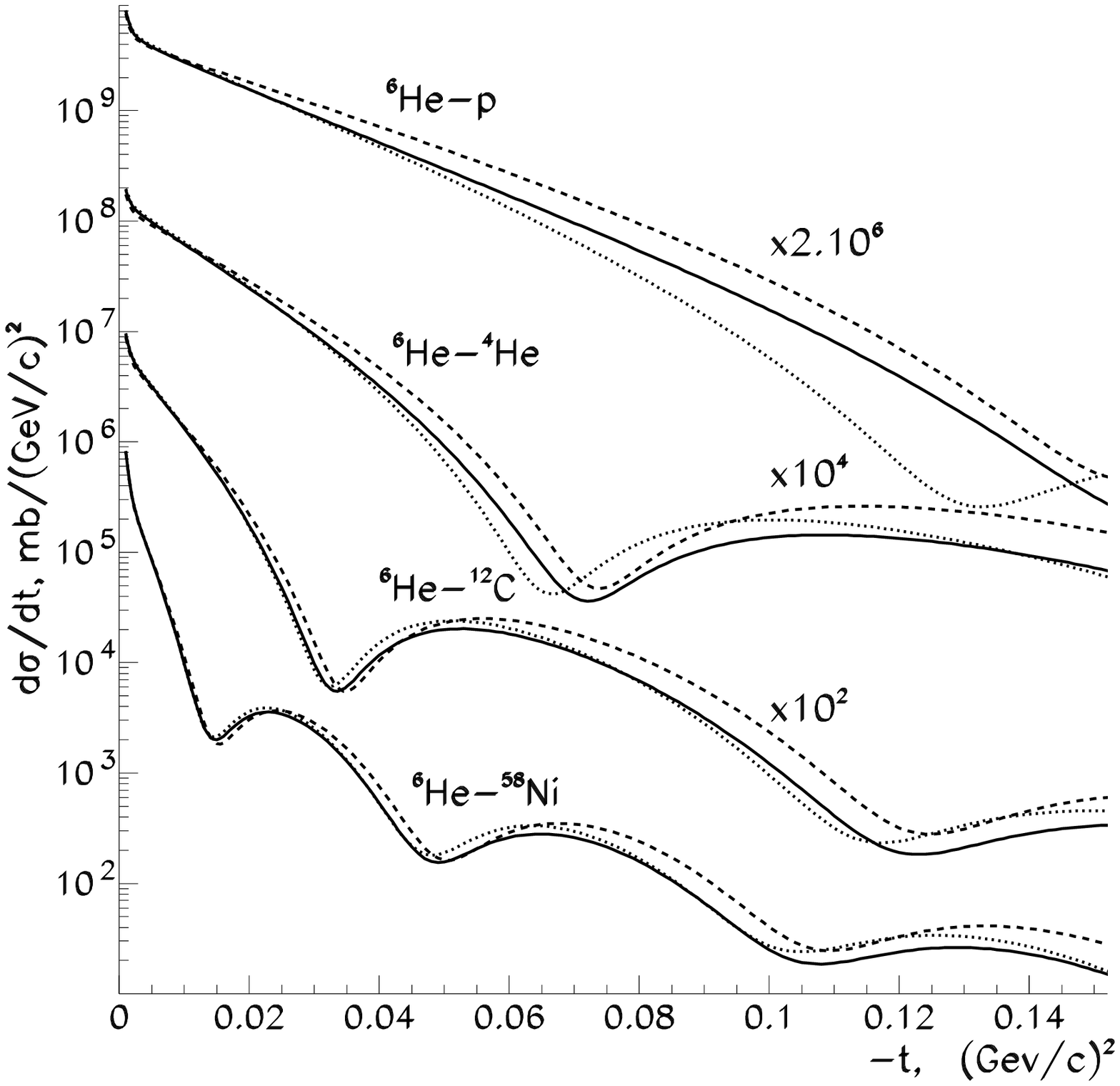,height=10cm} \caption{\small
The calculated cross sections for scattering of $^6$He on protons
and nuclear targets $^4$He, $^{12}$C and $^{58}$Ni at the energy of
0.8 GeV/nucleon as functions of the four-momentum transfer
squared~$- t$}
\end{figure}
= 2.88 fm, R$_{\rm m}$ = 2.30 fm (see Fig.~1). The dashed and dotted
curves correspond to density distributions without a halo (R$_{\rm
c}$ = R$_{\rm h}$), the dashed curve stands for R$_{\rm m}$ = 1.95
fm, and the dotted curve stands for R$_{\rm m}$ = 2.30~fm.

 It is seen that at small momentum transfers the behaviour of
the cross sections is governed by the value of R$_{\rm m}$ -- the
rms radius of the total nuclear matter distribution. Indeed, the
cross sections calculated for the matter density distributions of
different shape but with the same rms matter radius at $- t <$
0.03 (GeV/c)$^2$ are practically the same (compare the solid and
dotted curves in Fig. 4). At larger values of
$\mid\hspace{-1mm}t\hspace{-1mm}\mid$ the cross sections
corresponding to these two density distributions become
significantly different. A comparison of the calculated cross
sections for scattering of the $^6$He nuclei on different nuclear
targets shows that the sensitivity of the nucleus-nucleus cross
sections to the shape of the nuclear matter distribution in $^6$He
is not higher than that of the cross section for $^6$He-$p$
scattering.

To clear up the sensitivity of the cross sections to the nuclear
far periphery, we have calculated the relevant cross sections
using the nuclear matter densities in $^6$He with the density tail
and without it (Fig. 2). Since the tail contains a very small
amount of the total nuclear matter (on the level of 1\%), the
effect in the cross sections from the scattering on the nucleons
of the tail is so small that it is practically not seen in the
cross sections depicted in the logarithmic scale. Therefore, to
see  in the cross sections the effect of taking into account
this tail, we show in Fig. 5 the ratio of the cross sections
calculated for the matter density distributions with and without
the tail. We see that in the cross section for $^6$He-nucleus
scattering the size of the effect of taking into account the
$^6$He density tail is approximately the same as that in the cross
section for $^6$He-$p$ scattering.
Figure 5 also shows (by the dotted curves) the ratio of the cross
sections calculated for the $^6$He density without the tail with
the rms halo 
radii R$_{\rm h}$ = 2.88 fm and R$_{\rm h}$ = 3.34 fm (this halo
radius corresponds to the density version with the tail).
\begin{figure}[h]
\vspace{-1.45cm}
\centering \epsfig{file=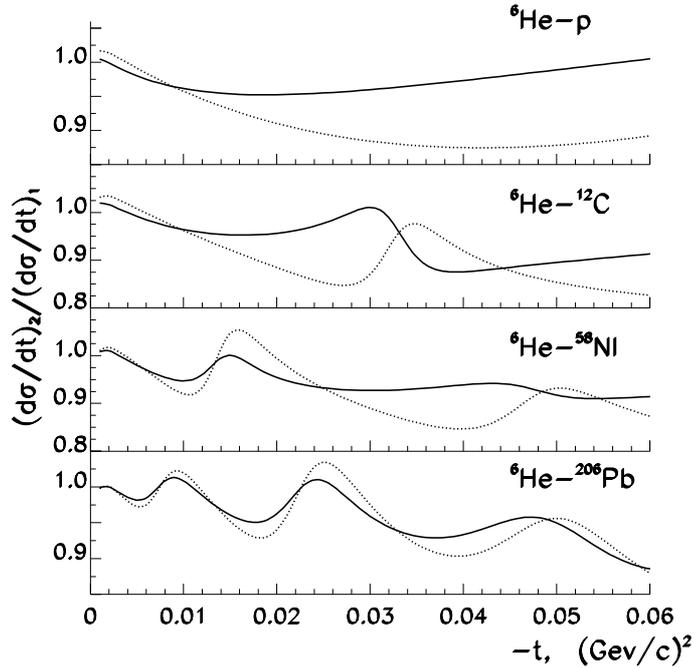,height=9.5cm}
\caption{\small The ratio of the cross sections (d$\sigma$/d$t)_2$
for scattering of $^6$He on protons and nuclear targets $^{12}$C,
$^{58}$Ni and $^{208}$Pb, calculated with the $^6$He matter
density distribution with a "tail" (see Fig. 2) to the
corresponding cross sections (d$\sigma$/d$t)_1$, calculated with
the distribution without this "tail" (solid curves). Dotted curves
stand for the ratios of the cross sections
(d$\sigma$/d$t)_2$/(d$\sigma$/d$t)_1$, calculated using the $^6$He
matter density distributions without a tail correspondingly for
the halo radii R$_{\rm h}$ = 3.34 fm and R$_{\rm h}$ = 2.88 fm}
\end{figure}

We see that in the cases of $^6$He-$p$ and $^6$He-$^{12}$C
scattering the effects in the cross sections due to taking into
account the density "tail" and due
to the corresponding increase of the halo radius in
the density without the "tail" are significantly different. However,
in the case of $^6$He scattering on heavy nuclei (here $^{208}$Pb)
the difference between the cross sections calculated with these two
density versions becomes smaller (for $^6$He-$^{208}$Pb scattering
the solid curve in Fig. 5 has the form similar to that of the dotted
curve). Therefore it would be difficult in this case to determine
whether the density distribution does have a "tail" or does not.
\begin{figure}[h]
\vspace{-1.0cm} \centering \epsfig{file=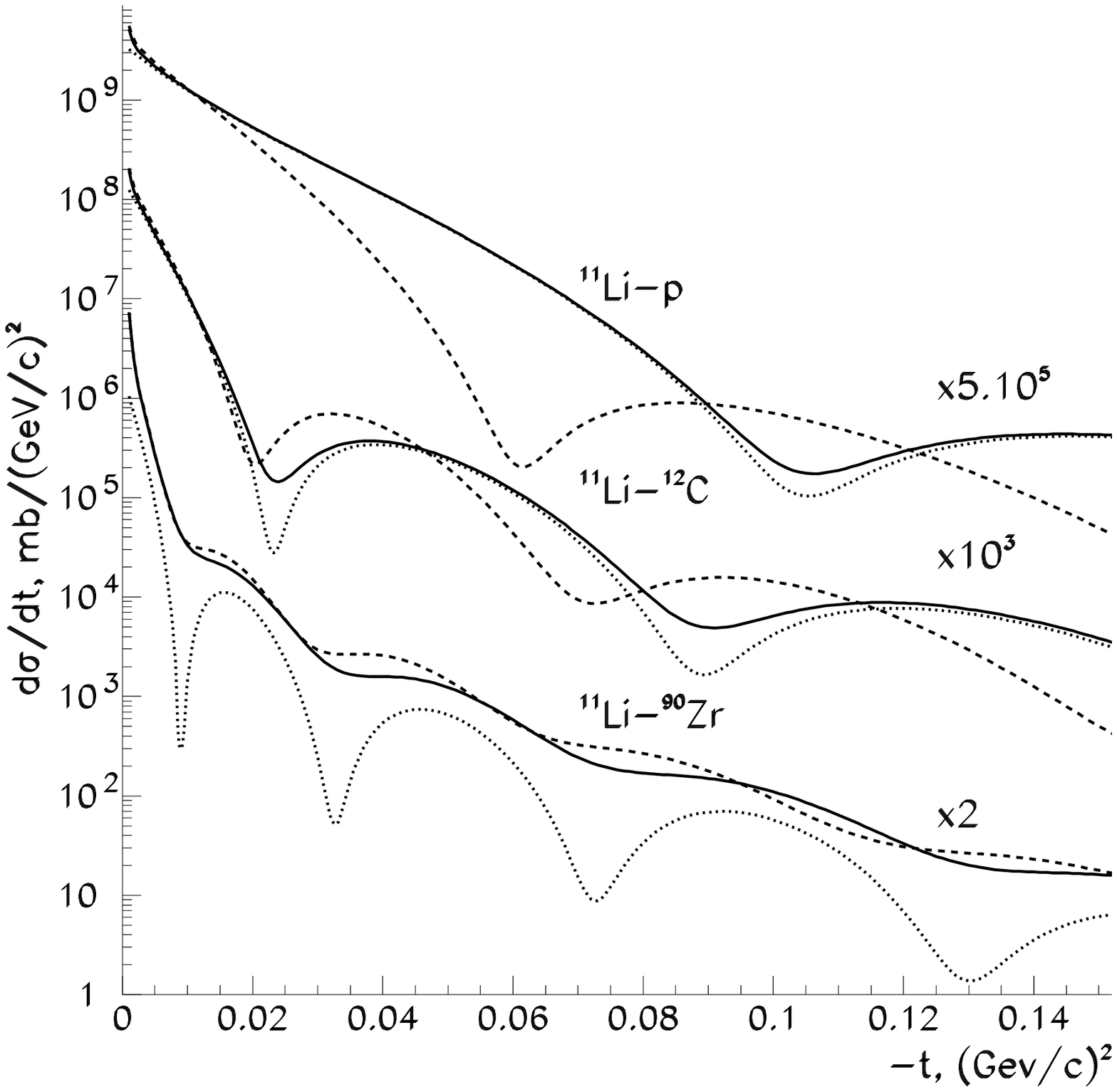,height=11cm}
\caption{\small The calculated cross sections for scattering of
$^{11}$Li on protons and nuclear targets $^{12}$C and $^{90}$Zr}
\end{figure}

Figure 6 shows the calculated cross sections for elastic
$^{11}$Li-$p$, $^{11}$Li-$^{12}$C and $^{11}$Li-$^{90}$Zr
scattering. The solid curves correspond to the calculations for the
$^{11}$Li nuclear matter density distribution with a halo (but
without a "tail"): R$_{\rm c}$ = \\2.50 fm, R$_{\rm h}$ = 5.86 fm,
R$_{\rm m}$ = 3.37 fm (see Fig. 3). The dashed curves correspond to
the calculations for the density distribution without a halo but
with the same rms matter radius: R$_{\rm c}$ =  R$_{\rm h}$ =
R$_{\rm m}$ = 3.37 fm. We see that the sensitivity of the cross
sections to the shape of the $^{11}$Li nuclear matter distribution
in the case when the target nucleus is heavy (here it is $^{90}$Zr)
is smaller than that in the case when the target is light (hydrogen
or $^{12}$C). Evidently, it is partly due to a significant
contribution to the cross section from the Coulomb scattering in the
case of heavy nuclear targets (targets with large Z). The dotted
curves in Fig. 6 show also the cross sections calculated for the
$^{11}$Li density distribution with a halo, but without taking into
account the contribution from the Coulomb scattering. We see that in
the $^{11}$Li-$p$ scattering cross section the contribution from the
Coulomb scattering  is small (this contribution is seen only at very
small values of $\mid\hspace{-1.34mm}t\hspace{-1.34mm}\mid < 0.005$
(GeV/c)$^2$ and in the region of the diffraction minimum). In the
$^{11}$Li-$^{12}$C cross section, the contribution from the Coulomb
scattering is more significant, while in the $^{11}$Li-$^{90}$Zr
cross section the contribution from the Coulomb scattering is
comparable to that of the strong interaction scattering, or even
surpasses it.

In order to clear up the sensitivity of the scattering cross
sections to the $^{11}$Li nuclear far periphery, we have
calculated the cross sections for $^{11}$Li scattering on several
nuclear targets for the $^{11}$Li matter density distributions
with and without the "tail". Figure 7 presents the results of the
calculations. We see that at small momentum transfers
$\mid\hspace{-1mm}t\hspace{-1mm}\mid$ the effect in the cross
sections from taking into account the "tail" of  the $^{11}$Li
matter distribution for the $^{11}$Li~-nucleus scattering is
approximately the same as that for $^{11}$Li-$p$ scattering. At
larger values of $\mid\hspace{-1mm}t\hspace{-1mm}\mid$, in the
regions of diffraction minima, the sensitivity of the cross
sections to the nuclear far periphery for $^{11}$Li-nucleus
scattering is somewhat higher than that for $^{11}$Li-proton
scattering.

 However, we should bear in mind that the calculated
cross sections in the diffraction minima are subject to noticeable
uncertainties (of the Coulomb contribution, of the Coulomb-nuclear
interference, due to some uncertainties in the ratio of the
real-to-imaginary parts of the NN-scattering amplitude and due to
some other reasons). 
Therefore, in the analyses of the measured
scattering cross sections with the purpose to study nuclear matter
distributions it is desirable not to use the cross sections in the
regions of the diffraction minima.
\begin{figure}[h]
\vspace{-0.5cm}
\centering \epsfig{file=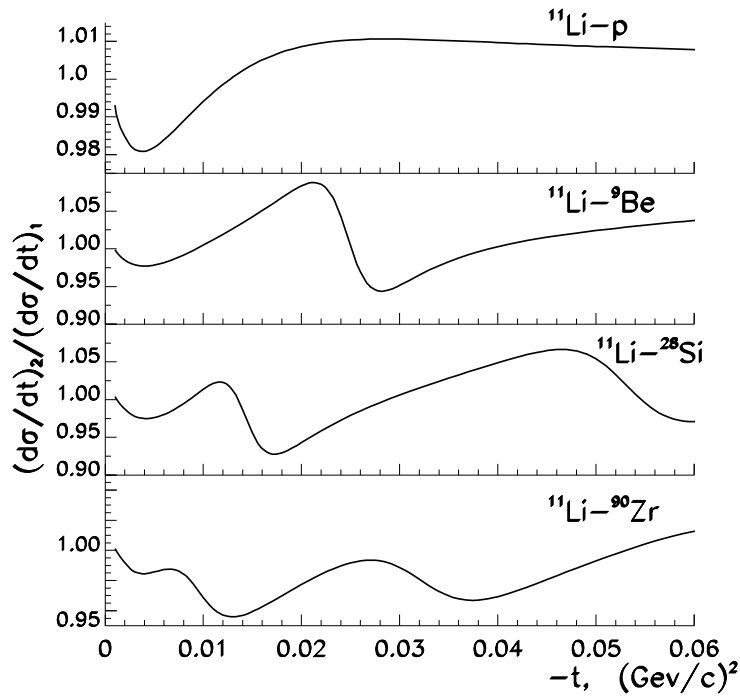,height=10cm}
\caption{\small The ratio of the cross sections
(d$\sigma$/d$t$)$_2$ for scattering of the $^{11}$Li nuclei on
hydrogen and nuclear targets $^9$Be, $^{28}$Si and $^{90}$Zr,
calculated with the $^{11}$Li matter density with a "tail" (Fig.
3), to the corresponding cross sections (d$\sigma$/d$t$)$_1$,
calculated without the "tail"}
\end{figure}

\section{Conclusion}
The whole set of the carried out considerations gives evidence
that the sensitivity of the cross sections for nucleus-nucleus
scattering to the shape of the nuclear matter distribution of the
studied exotic nuclei is of the same order of magnitude as that of
the cross sections for nucleus-proton scattering. We do not see
marked advantages in using cross sections for elastic
nucleus-nucleus scattering with the aim to study nuclear matter
distributions in exotic nuclei as compared to using cross sections
for nucleus-proton scattering. It should be noted also that both
the measurements of the differential cross
sections and calculations of the cross
sections with the exact Glauber formula for
nucleus-nucleus scattering are noticeably more complicated tasks
than the corresponding measurements and calculations for
nucleus-proton scattering. We come to the conclusion that
performing experiments to measure differential cross sections for
elastic nucleus-nucleus scattering with the purpose to study
nuclear matter distributions in exotic nuclei is hardly justified.

\end{document}